\documentclass[aps,prd,preprint,preprintnumbers,unsortedaddress,superscriptaddress,showpacs,nofootinbib]{revtex4-1}
\pdfoutput=1
\usepackage{graphicx}
\usepackage{amsmath}
\usepackage{amsfonts}
\usepackage{amssymb}
\usepackage{color}
\usepackage{ulem}

\usepackage[colorlinks, citecolor=blue,anchorcolor=red,menucolor=red, linkcolor=red,filecolor=red,runcolor=red,urlcolor=blue,frenchlinks=red]{hyperref}

\usepackage{bm}
\usepackage[section]{placeins}
\usepackage{booktabs}

\begin{document}
\arraycolsep1.5pt

\title{$J/\psi$ decay into $\phi (\omega)$ and  vector-vector molecular states}

\author{R. Molina}
\affiliation{ School of Physics, Beihang University, Beijing 100191, China}
\affiliation{
Departamento de Fisica Teorica II, Plaza Ciencias, 1, 28040 Madrid, Spain}

\author{L.~R.~Dai}
\affiliation{Department of Physics, Liaoning Normal University, Dalian 116029, China}

\author{L. S. Geng}
\affiliation{ School of Physics \& Beijing Advanced Innovation Center for Big Data-based Precision Medicine, Beihang University, Beijing100191, China}
\affiliation{School of Physics, Zhengzhou University, Zhengzhou, Henan 450001, China}

\author{E. Oset}
\affiliation{ School of Physics, Beihang University, Beijing 100191, China}
\affiliation{Departamento de F\'isica Te\'orica and IFIC, Centro Mixto Universidad de Valencia-CSIC,
Institutos de Investigaci\'{o}n de Paterna, Aptdo. 22085, 46071 Valencia, Spain
}

\date{\today}
\begin{abstract}
Based on the picture that the $ f_0(1370), f_0(1710), f_2(1270), f'_2(1525), \bar{K}^{*0}_2(1430)$ resonances are dynamically generated from the vector-vector interaction, we study the decays $J/\psi \to \phi (\omega) f_0(1370) [f_0(1710)]$,  $J/\psi \to \phi (\omega) f_2(1270) [f'_2(1525)]$,  and $J/\psi \to K^{*0} \bar{K}^{*0}_2(1430)$ and make predictions for seven independent ratios that can be done among them. The starting mechanism is that the $J/\psi$ decays into three vectors, followed by the final state interaction of a pair of them. The weights of the different three vector primary channels are obtained from the basic assumption that the $J/\psi$ ($c \bar c$) is an SU(3) singlet.  By means of only one free parameter we predict four ratios in good agreement with experiment, make two extra predictions for rates yet unmeasured, and disagree on one data for which only upper bounds are reported. Further measurements are most welcome to complete the information required for these ratios that are testing the nature of these resonances as dynamically generated. 
 \end{abstract}

\maketitle


\section{Introduction}
The undeniable success of the quark model to put in order the bulk of hadronic states \cite{isgur, karl, capstick,  roberts, vijande} has not precluded that some states show a richer structure than the
standard  $q\bar{q}$  meson and $qqq$ baryon composition. Many different structures have been proposed to understand different hadronic states, as tetraquarks, pentaquarks, hybrids, molecules (see recent
reviews on these topics \cite{chen,lebed, pilloni, guo, stone, olsen, karliner, liu,hanhartxyz}). The molecular picture to describe  many hadronic states \cite{guo} received an undeniable boost, with  a
broad consensus reached that the recently observed pentaquark states \cite{pentaex} are of molecular
 nature \cite{pentaex,zhusl,pavon,jhe,zhifeng,xiao,zhang,meng,voloshin, elena, zgwang, manapavon, xu, manalogeng, burns, zou, hosaka}, as originally predicted in Ref. \cite{molinawu}.
 
 It is fair to recall that there are also many such cases of  molecular nature in the light sector. The  case of light scalar mesons, $f_0(500)$, $f_0(980)$, $a_0(980)$, as dynamically generated  resonances from the
 pseudoscalar-pseudoscalar interaction \cite{npa,kaiser, markushin, juan} has received much attention (see reviews \cite{angelsoller,review}). So is the case of axial vector mesons, generated from the
  interaction of pseudoscalar and vector mesons \cite{lutz,roca},  or particular  baryons  like the two $\Lambda(1405)$,
  $N^*(1535)$, $\cdots $, generated from the meson-baryon   interaction \cite{weise,ramos, ollerulf, kolo, inoue, carmen, jido, hyodo, kamiya}.

  With so much work done in these sectors,  it is surprising that the interaction of vector mesons among themselves  has received comparatively much less attention.
  The reason probably  is that  the chiral Lagrangians  \cite{weinberg,gasser} do not deal with the interaction of vector mesons among  themselves  and the works reported  before
  rely upon the  chiral Lagrangians using a unitary extension that matches them at low energies.

The link of the former works to the vector-vector interaction can be traced to the work of Ref. \cite{rafael}, where it was shown that the chiral Lagrangians up to $O(p^4)$, and implementing vector meson
dominance, can be equally  obtained using the local hidden gauge approach \cite{hidden1,hidden2,hidden4,hideko} through the exchange of vector mesons. The local hidden gauge approach,
dealing explicitly with vector mesons, provides the tools  for the vector-vector interaction through a contact term and the exchange of vector mesons.  This formalism was used in \cite{raquel,gengvec}
to study first the $\rho\rho$ interaction \cite{raquel} and then extended to SU(3) \cite{gengvec}. As a consequence of that, vector-vector bound states that decay into lighter mesons were found in these
works, and the  $f_0(1370)$, $f_2(1270)$, $f_0(1710)$, $f'_2(1525)$, $\bar{K}^{*0}_2(1430)$ resonances, among others, were generated within this approach.  The consistency of this picture with many reactions where
these resonances are produced has been tested, for instance, in $\bar{B}^0$ and $\bar{B}^0_s$ decays \cite{xie}, photoproduction of the $f_{2}(1270)$,  $f'_2(1525)$ \cite{xiephoto,xieogeng}, the coupling of
$f_0(1370)$ and  $f_2(1270)$ to $\gamma\gamma$ \cite{yamagata},  the $J/\psi$ decay into $\phi (\omega)$ and some of these  resonances, together with decay  to $K^{*0}$ and the $\bar{K}^{*0}_2 (1430)$ \cite{alberdai}, and the
production  of these resonances in $\psi (nS)$ and $\Upsilon (nS)$ decays \cite{dai,dai2}.

In the present work we want to retake the issue of the $J/\psi$ decay \cite{alberdai}. In that work we introduced some Lagrangians, involving transition of  $J/\psi$ to three vectors  by analogy to a
similar work in $J/\psi \to  \phi(\omega) PP$,  with $P$ a pseudoscalar meson, where the PP mesons interact producing the $f_0(500)$ and  $f_0(980)$ resonances  \cite{ollerpsi,palomar,lahde}. In  Refs. \cite{ollerpsi} and
\cite{palomar} two different formalisms, although equivalent, were used to study the process. In between,  different methods, based on the $c\bar{c}$ character as a singlet of SU(3) have been developed, which
are conceptually easier. Indeed, in Ref. \cite{liangsakai}, where the  $J/\psi \to \eta (\eta') h_1(1380)$ BESIII reaction has been studied \cite{besh1}, the basic ingredient  is the $J/\psi \to VVP$ ($V$ for vector)
transition, followed by $VP$ interaction that, according to \cite{lutz,roca} generates the axial  vector mesons. The $J/\psi$ is assumed  to be an SU(3) singlet and then  two structures  are used $\langle VVP \rangle$
and $\langle VV \rangle \langle P \rangle$, with  $V$ and $P$  the SU(3) matrices of vectors and  pseudoscalars in SU(3), and  $\langle  \rangle$ standing for the trace of these matrices.  The formalism of Ref. \cite{liangsakai}
is shown to be equivalent to those of Refs. \cite{ollerpsi,palomar}, yet technically easier, and more intuitive, with the dominant $\langle VVP \rangle$ contribution found to be equivalent to the main term in Refs.
\cite{ollerpsi} and \cite{palomar}.

The  dominance of the trace of three mesons has also been established in the $\chi_{c1}$ decay into $\eta\pi^+\pi^-$ \cite{karmicer}, where the  $\langle PPP \rangle$  structure for primary production
of three pseudoscalars in the  $\chi_{c1}$ decay is followed by the interaction of $PP$ pairs to produce the $a_0(980)$ and  the $f_0(500)$ resonances \cite{liangchi1}.  Similarly, the $\langle VVP \rangle$
structure has also been tested  as the dominant one in the study of the $\chi_{cJ}$ decay into $\phi h_1(1380)$ \cite{besphih1}, where $\phi$ is a spectator  and the other $VP$ pair interacts to produce the
$h_1(1380)$ \cite{shengyuan}.

In the present work we want to study the $J/\psi$ decay into $\phi (\omega)$ and  vector resonances that come from the vector-vector interaction. We then assume that the $J/\psi (c\bar{c})$  is an SU(3) singlet
and take the $\langle VVV \rangle$ primary production, followed by the interaction of the $VV$ pairs.  We shall also consider  a possible small contribution of the  $\langle VV \rangle \langle V \rangle$
structure, which has some similarity to the $\langle VV \rangle \langle P\rangle$ structure of \cite{liangsakai}, but cannot be taken from there since these correspond to different structures.

The issue of the vector-vector interactions  has been vindicated recently after the debate  created by the works  \cite{gulmez, guodis}  questioning the $f_{2}(1270)$
as a $\rho\rho$ molecular state. The claims of these works contradict basic Quantum Mechanics rules because the $f_0(1370)$ state was produced in Refs. \cite{gulmez,guodis}, as in Refs. \cite{raquel,gengvec}, and the potential in the $J=2$ sector
is more than twice bigger than for $J=0$ and attractive, hence the $f_2$ state  should be more bound than the $f_0$, as is the case in Refs.  \cite{raquel,gengvec}. Without entering the discussion here, the steps
that invalidated the approaches of Refs. \cite{gulmez} and \cite{guodis} were disclosed  in Refs. \cite{contra1} and \cite{contra2}, respectively, and at the same time an improved method was proposed which, after small refitting
in the cutoff used to regularize the loops, produced the $f_{2}(1270)$ as a bound $\rho\rho$ state, and very importantly,  produced couplings of the $f_2$ to $\rho\rho$ practically identical to those found in Refs.  \cite{raquel,gengvec}.
With this reassurance, we shall continue to use the couplings and $G$ functions of Refs. \cite{raquel,gengvec} which are needed as input for the calculations that we perform here.

Apart from the new perspective and  formalism discussed above, there is one more reason to retake this problem since new data are available. Then we shall  evaluate  the rate  for 
$J/\psi \to \phi (\omega) f_0(1370) [f_0(1710)]$,  $J/\psi \to \phi (\omega) f_2(1270) [f'_2(1525)]$,  and $J/\psi \to K^{*0} \bar{K}^{*0}_2(1430)$ and from  there we shall evaluate ratios that we will
compare with experiment.

This paper is organized as follows. In Sec. II, we explain how the $J/\psi$ first decays into three vector mesons and then a pair of them rescatter to generate dynamically the $f_0(1370)$,  $f_0(1710)$, $f_2(1270)$, $f'_2(1525)$, and $\bar{K}^{*0}_2(1430)$. We then construct seven ratios to eliminate unknown couplings and fix one remaining relative weight by fitting to four experimental known ratios and make predictions for three remaining ones in Sec. III. Followed by a short summary and outlook in Sec. IV. 

\section{Formalism}
We study first the $J/\psi \to VVV$ transition for which we will have the operator $ {\cal C}\langle VVV \rangle$  mixed with  $ C \beta \langle VV \rangle \langle V \rangle$, with
$C$ a normalization constant  and $\beta$ a parameter  to fit to the data. The matrix containing the nonet of vector mesons can be written as,
\[
V_\mu=
  \begin{pmatrix}
    \frac{1}{\sqrt{2}}\rho^0
+\frac{1}{\sqrt{2}}\omega & \rho^+ & K^{*+}  \\
   \rho^-& -\frac{1}{\sqrt{2}}\rho^0+\frac{1}{\sqrt{2}}\omega & K^{*0}\\
   K^{*-}&\bar{K}^{*0}&\phi
  \end{pmatrix}_\mu
\]
Then, we can construct the trace, $\langle V \cdot VV \rangle$, and look at the terms which accompany the $\omega$, $\phi$ and $K^{*0}$ mesons. 
We find,
\begin{itemize}
  \item[1)] for $\omega$ meson
 \begin{eqnarray}  \label{eq:m1}
 \frac{\omega}{\sqrt{2}} \big\{ 3 \rho^0 \rho^0 + 3 \rho^+ \rho^- + 3 \rho^- \rho^+ + \omega\omega + 3  K^{*+}  K^{*-} + 3 K^{*0} \bar{K}^{*0} \big\} \,,
\end{eqnarray}
   \item[2)] for $\phi$  meson
  \begin{eqnarray} \label{eq:m2}
  \phi \big\{ 3  K^{*+}  K^{*-} +  3 K^{*0} \bar{K}^{*0} +\phi\phi \big\}  \,,
  \end{eqnarray}
   \item[3)] for  $K^{*0}$   meson
     \begin{eqnarray}\label{eq:m3}
K^{*0} \big\{ 3 \rho^+   K^{*-} -  \frac{3}{\sqrt{2}}  \rho^0   \bar{K}^{*0} + \frac{3}{\sqrt{2}}\omega \bar{K}^{*0} + 3 \phi \bar{K}^{*0} \big\} \,.
  \end{eqnarray}
\end{itemize}

It is convenient to write these in terms of isospin states in order to use the information of Ref. \cite{gengvec}.  We have the isospin
multiplets $(-\rho^+,\rho^0,\rho^-)$,
$(K^{*+},{K}^{*0})$, $(\bar{K}^{*0},-K^{*-})$  and then
\begin{eqnarray} \label{eq:im1}
&|\rho\rho, I=0\rangle = -\frac{1}{\sqrt{6}} |\rho^0 \rho^0 + \rho^+ \rho^- + \rho^- \rho^+ \rangle  \,;  \nonumber \\
&|\omega \omega, I=0\rangle =\frac{1}{\sqrt{2}} |\omega \omega \rangle \,; \quad  |\phi\phi, I=0\rangle =\frac{1}{\sqrt{2}} |\phi\phi \rangle \,,
\end{eqnarray}
where we introduced the extra $1/\sqrt{2}$ factor of the unitary normalization for identical particles,
\begin{eqnarray}\label{eq:im2}
|K^* \bar{K}^*, I=0\rangle =-\frac{1}{\sqrt{2}} |K^{*+}K^{*-} + K^{*0}\bar{K}^{*0} \rangle  \,,
\end{eqnarray}
\begin{eqnarray}\label{eq:im3}
|\rho K^*, I=1/2, I_3=1/2 \rangle =\sqrt{\frac{2}{3}} | \rho^+ K^{*-}\rangle  -\sqrt{\frac{1}{3}} | \rho^0 \bar{K}^{*0} \rangle  \,.
\end{eqnarray}
From Eqs.(\ref{eq:m1}-\ref{eq:m3}) and (\ref{eq:im1}-\ref{eq:im3}) we find the weight for $|\rho\rho, I=0\rangle$, $|K^* \bar{K}^*, I=0\rangle$,  $\omega\omega$
and $\phi\phi$ production  in isospin basis as
\begin{eqnarray}\label{eq:h}
&h_{\omega \rho \rho}=-\frac{3\sqrt{3}}{2} \,; \quad h_{\omega K^* \bar{K}^*}=-3    \,; \quad   h_{\omega \omega \omega}=3   \,; \quad  h_{\omega \phi\phi}=0    \,,\nonumber \\
&h_{\phi \rho \rho}=0 \,; \quad h_{\phi K^* \bar{K}^*}=-3 \sqrt{2} \,; \quad   h_{\phi \phi\phi}= 3 \sqrt{2} \,; \quad    h_{ \phi \omega \omega}=0  \,, \nonumber \\
&h_{K^{*0} \rho \bar{K}^*}=3 \sqrt{\frac{3}{2}} \,; \quad h_{K^{*0} \omega \bar{K}^*}= \frac{3}{\sqrt{2}} \,;  \quad h_{K^{*0} \phi \bar{K}^*}=3 \,,
\end{eqnarray}
where the symmetry factor $n!$ for production of $n$ identical particles has been taken into account.

\subsection{Terms with $\langle VV \rangle \langle V \rangle$}
Alternatively, we can see what weight we get with the $\langle VV \rangle \langle V \rangle$  structure.  In this case we find that
\begin{eqnarray}\label{eq:VVcV}
\langle VV \rangle \langle V \rangle \equiv (\sqrt{2} \omega +\phi) \big\{ \rho^0 \rho^0 + \rho^+ \rho^- + \rho^- \rho^+
+ \omega\omega + 2 K^{*+}K^{*-} + 2  K^{*0}\bar{K}^{*0}  +\phi\phi \big\} \,,
\end{eqnarray}
and  the weights are now
\begin{eqnarray}\label{eq:hp}
&h'_{\omega \rho \rho}=-\sqrt{3} \,; \quad h'_{\omega K^* \bar{K}^*}=-4    \,; \quad   h'_{\omega \omega \omega}=6   \,; \quad  h'_{\omega \phi\phi}=2    \,,\nonumber \\
&h'_{\phi \rho \rho}=-\sqrt{\frac{3}{2}} \,; \quad h'_{\phi K^* \bar{K}^*}=-2 \sqrt{2} \,; \quad   h'_{\phi \phi\phi}= 3\sqrt{2} \,; \quad    h'_{ \phi \omega \omega}=\sqrt{2}  \,, \nonumber \\
&h'_{K^{*0} \rho \bar{K}^*}=0 \,; \quad h'_{K^{*0} \omega \bar{K}^*}= 2\sqrt{2} \,;  \quad h'_{K^{*0} \phi \bar{K}^*}=2 \,.
\end{eqnarray}
\subsection{Production of resonances }
\begin{figure}[ht!]
\includegraphics[scale=0.7]{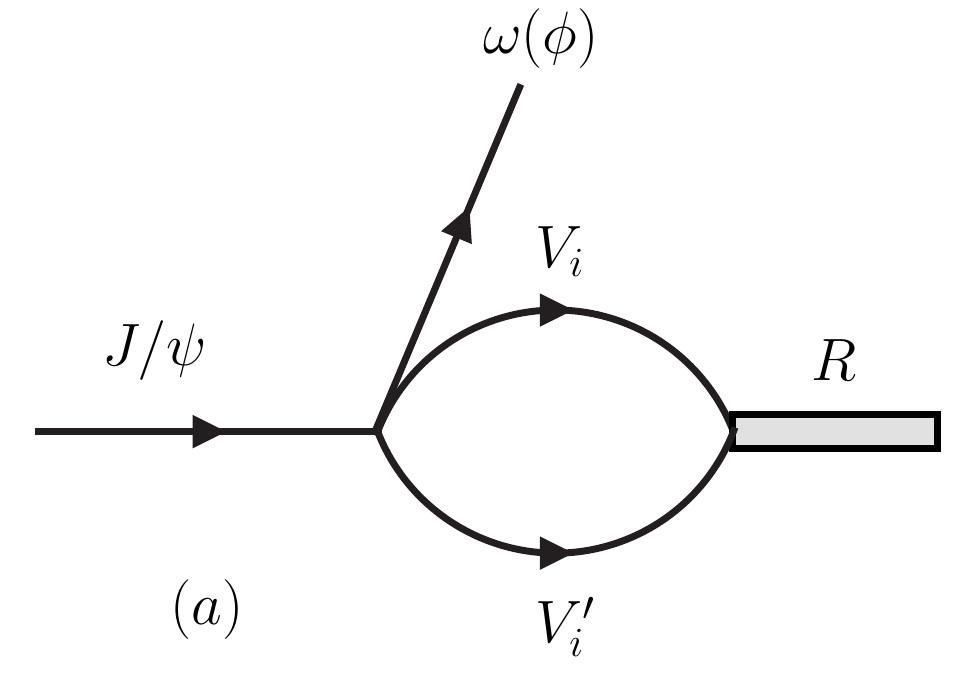}~~~~\includegraphics[scale=0.7]{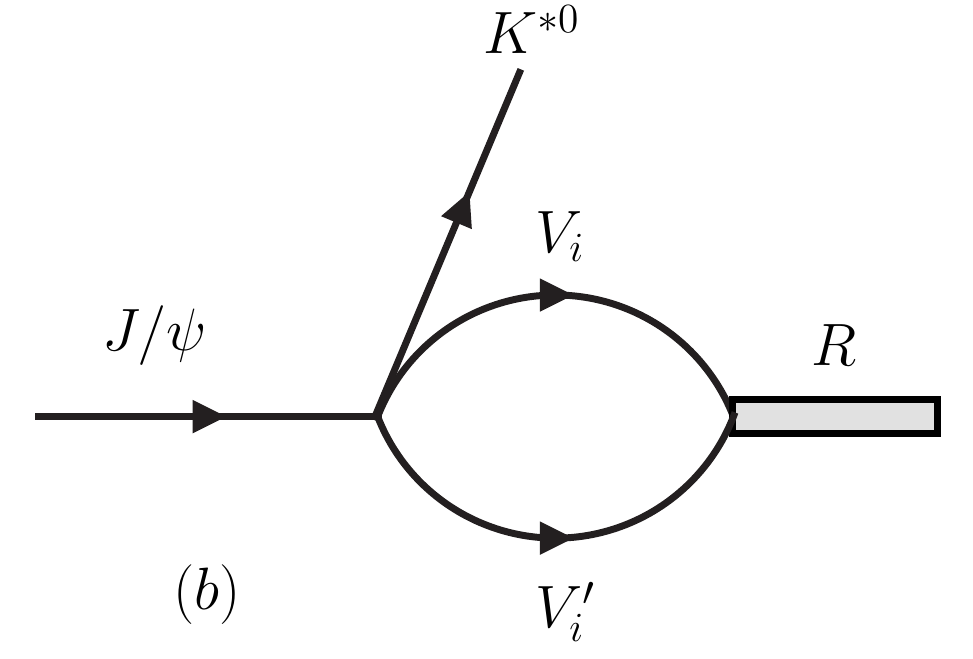}
\caption{Mechanisms to produce the resonance $R$ together with a $\phi~(\omega)$ (a) or a  $K^{*0}$ (b).}
\label{figa}
\end{figure}

The production of the resonances proceeds as described graphically  in Fig. \ref{figa}

Analytically we can translate the mechanism of Fig. \ref{figa} into the transition matrix
\begin{eqnarray}\label{eq:t}
t_j={\cal C} \sum_{i} h_i \, G_i(M_{R_j}) \, g_{R_j,i} \,,
\end{eqnarray}
where $h_i$ are the weights calculated before for the different intermediate channels, $G_i$ is the loop function of the two intermediate mesons
and  $g_{R_j,i}$ are the couplings of the resonance $g_{R_j}$  to these channels $i$. This information  we obtain from Ref. \cite{gengvec} and  is shown in  Table  \ref{coupGs},
together with the same uncertainties in Table \ref{coupG} that were evaluated in \cite{alberdai}.

\begin{table}[h!]
\caption{Couplings and values of the loop functions for the different channels at the resonance energy for the $f_{0}(1370)$ and $f_{0}(1710)$ states.}
\vspace{0.3cm}
\centering
\begin{small}
\begin{tabular}{l c| c cccc}
\hline\hline
 &&  &$\rho\rho$&$K^* \bar{K}^*$&$\omega\omega$&$\phi\phi$\\
[0.1ex]
\hline
 & &$g_{i} (\textrm{MeV})$ & ~$7914 - i 1048$ ~&~ $1210-i415$~ & ~$-39+i31$~ &~ $12+i24$ \\[1ex]
& & error $g_{i} (\%) $
& $4$ &$3$& $22$ & $22$ \\[0.1ex]
\raisebox{1.5ex}{}  &\raisebox{1.5ex}{$f_{0}(1370)$} &$G_{i} (\times 10^{-3})$
&  $-7.69+i1.72$ & $-4.13+i0.26$ & $-8.97+i0.87$ & $-0.63+i0.14$ \\[0.1ex]
& &error $G_{i} (\%)$
&$10$  &$29$&$42$  &$220$  \\[0.1ex]
 \hline
& &$g_{i} (\textrm{MeV})$ & $-1030+i1087 $ & $7127+i94$ & $-1764+i109$ & $-2494+i205$   \\[1ex]
 & & error $g_{i} (\%) $
& $12$ &$6$&$2$  & $2$  \\[0.1ex]
\raisebox{1.5ex}{} & \raisebox{1.5ex}{$f_{0}(1710)$}& $G_{i} (\times 10^{-3})$
&$-9.7+i6.18$ & $-7.68+i0.58$ & $-10.85+i8.19$ & $-2.16+i0.13$ \\[1ex]
& &error $G_{i} (\%)$
&$6$  &$17$&$19$  &$141$   \\[0.1ex]
\hline\hline
\end{tabular}
\end{small}
\label{coupGs}
\end{table}

\begin{table}[h!]
\caption{Couplings and values of the loop functions for the different channels at the resonance energy for the $f_{2}(1270)$, $f'_{2}(1525)$ and ${\bar{K}}_{2}^{*\,0}(1430)$ states.}
\vspace{0.3cm}
\centering
\begin{small}
\begin{tabular}{l c| c ccccccc}
\hline\hline
 &&  &~~~~$\rho\rho$~~~~&~~~~$K^* \bar{K}^*$ ~~~~ &~~~~ $\omega\omega$ ~~~~&$\phi\phi$ ~~~~&~~~~ $\rho\bar{K}^*$~~ ~~&~~~~$\omega\bar{K}^{*\,0}$~~~~&~~~~$\phi\bar{K}^{*\,0}$~~~~\\
[0.1ex]
\hline
 & &$g_{i} (\textrm{MeV})$ &10551 & $4771$ & $-503$ & $-771$&0&0&0   \\[1ex]
 & & error $g_{i} (\%) $
&  $4$ & $3$ & $22$ & 22&0&0&0  \\[0.1ex]
\raisebox{1.5ex}{}  &\raisebox{1.5ex}{$f_{2}(1270)$} &$G_{i} (\times 10^{-3})$
&  $-4.74$ & $-3.00$ & $-4.97$ & 0.475&0&0&0  \\[0.1ex]
& &error $G_{i} (\%)$
&  $10$ & $29$ & $42$ & 220&0&0&0  \\[0.1ex]
 \hline
& &$g_{i} (\textrm{MeV})$ & $-2611$ & $9692$ & $-2707$ & $-4611$&0&0&0   \\[1ex]
 & & error $g_{i} (\%) $
&  $12$ & $6$ & $2$ & 2&0&0&0  \\[0.1ex]
\raisebox{1.5ex}{} & \raisebox{1.5ex}{$f^\prime_{2}(1525)$}& $G_{i} (\times 10^{-3})$
&$-8.67$ & $-4.98$ & $-9.63$ & $-0.710$&0&0&0 \\[1ex]
& &error $G_{i} (\%)$
&  $6$ & $17$ & $19$ & 141&0&0&0  \\[0.1ex]
\hline
& &$g_{i} (\textrm{MeV})$ & 0 & $0$ & 0 & 0&10613&2273&$-2906$   \\[1ex]
 & & error $g_{i} (\%) $
&  $0$ & $0$ & $0$ & 0&3&5&5  \\[0.1ex]
\raisebox{1.5ex}{} & \raisebox{1.5ex}{${\bar{K}}_{2}^{*\,0}(1430)$}& $G_{i} (\times 10^{-3})$
&0& $0$ & 0 & 0&$-6.41$&$-5.94$&$-2.70$ \\[1ex]
& &error $G_{i} (\%)$
&  $0$ & $0$ & $0$ & 0&12&19&43  \\[0.1ex]
\hline\hline
\end{tabular}
\end{small}
\label{coupG}
\end{table}

If we consider the $\langle VV \rangle \langle V \rangle$ structure, then we replace
\begin{eqnarray}
h_i  \rightarrow h_i + \beta h'_i\,.\label{eq:beta}
\end{eqnarray}
We have ignored the spin  structure  with the four vectors that we have in the production vertex. One can easily implement it using  the spin operators  of Refs. \cite{raquel,gengvec}.
However, it is unnecessary since  we only calculate  ratios between spin
$J = 2$ states and $J=0$  states independently, and factors coming from the spin structure  cancel in the ratios.    
 
\section{Results and discussions}

For the evaluation of the $J/\psi$ decay into $\omega(\phi)$ and a molecular VV state, the standard formula for the width is given by
\begin{equation}\label{eq:ga}
\Gamma=\frac{1}{8\pi}\frac{1}{M^2_{J/\psi}}\vert t\vert^2q\ ,
\end{equation}
with $q=\lambda^{1/2}(M^2_{J/\psi},M^2_R, M^2_V)/2M_{J/\psi}$, where $M_R$ stands for the mass of the VV resonance, and $M_V$ refers to the mass of the vector meson, $\omega, \phi$ or $K^{*0}$, involved in the $J/\psi$ decay. For the evaluation of $\Gamma$ and $q$ in Eq. (\ref{eq:ga}), we take the values of the masses of the resonances given in Ref. \cite{gengvec}, although, if the nominal masses of the resonances according to the PDG \cite{pdg} are used instead, numerical results barely change. The ratios,
\begin{eqnarray}
 &&R_1\equiv\frac{\Gamma_{J/\psi\to\phi f_2(1270)}}{\Gamma_{J/\psi\to\phi f'_2(1525)}}\quad R_2\equiv\frac{\Gamma_{J/\psi\to\omega f_2(1270)}}{\Gamma_{J/\psi\to\omega f'_2(1525)}}\nonumber\\
 &&R_3\equiv\frac{\Gamma_{J/\psi\to\omega f_2(1270)}}{\Gamma_{J/\psi\to\phi f_2(1270)}}\quad R_4\equiv\frac{\Gamma_{J/\psi\to K^{*0}\bar{K}^{*0}_2(1430)}}{\Gamma_{J/\psi\to\omega f_2(1270)}}\ .\label{eq:ra1}
\end{eqnarray}
are estimated as it was done in Ref. \cite{alberdai}, but with the new weights given in Eqs. (\ref{eq:h}) and (\ref{eq:hp}), by using Eqs. (\ref{eq:t}), (\ref{eq:beta}) and (\ref{eq:ga}). In addition, we evaluate here three more ratios:
\begin{eqnarray}
R_5\equiv\frac{\Gamma_{J/\psi\to\omega f_0(1370)}}{\Gamma_{J/\psi\to\omega f_0(1710)}}\quad R_6\equiv\frac{\Gamma_{J/\psi\to \phi f_{0}(1370)}}{\Gamma_{J/\psi\to\phi f_0(1710)}}\quad R_7\equiv\frac{\Gamma_{J/\psi\to\omega f_0(1710)}}{\Gamma_{J/\psi\to\phi f_0(1710)}}\ .\label{eq:ra2}
\end{eqnarray}
First of all, the parameter $\beta$ in Eq. (\ref{eq:beta}) is fitted to get the experimental values of the ratios of Eqs. (\ref{eq:ra1}) and (\ref{eq:ra2}). These values are gathered in the last column of Table \ref{tab:ra}. In the PDG \cite{pdg}, and for the $J/\psi$ decay, we find experimental measurements for the decays involved in the ratios $R_i$, $i=1,3,4,7$. While for $R_2$, there is only an upper limit quoted  for $\Gamma(J/\psi\to \omega f'_2(1525))$ of $0.2\times 10^{-3}$, which has about $50$\% uncertainty. For this reason, we include in the fit experimental data on the four ratios for $i=1,3,4,7$, while those for $i=2,5,6$ become predictions of the theory.  We obtain, $\beta=0.32$, with the values of the ratios given in the fourth column of Table \ref{tab:ra}. The fit passes the Pearson's $\chi^2$ test at a $90$\% upper confidence limit. Then, $95$\% confidence intervals (CI) are estimated using two different methods:
(I) The parameter $\beta$ is kept fixed, and the ratios are evaluated by generating random numbers of $g_i$, $G_i$ assuming those to be normally distributed with the mean and standard deviation given by the central values and errors of these quantities as shown in Tables \ref{coupGs} and \ref{coupG}; (II) With bootstrap, by generating a new set of experimental data of these ratios by assuming these are normally distributed, as well as for the input variables, $g_i$ and $G_i$, and readjusting the $\beta$ parameter. From this method, the error of the $\beta$ parameter is estimated as the standard deviation obtained. We get $\beta=0.3\pm 0.3$. In both cases, (I) and (II), the size of the generated random sample is $n=1000$. Results are given in Table \ref{tab:ra}.
\begin{table}[h!]
\caption{Comparison of the $95$\% CI estimated by (I) fixing $\beta$ and (II) readjusting $\beta$ from a new random data sample as explained in the text. }
\vspace{0.3cm}
 \begin{center}
  \begin{tabular}{ccccccc}
  \hline\hline
  Ratio&$\beta=0$&$\beta=1$&$\beta=0.32$&$CI_{95\%}^{(I)}$&$CI_{95\%}^{(II)}$&Exp.\\
  [0.1ex]
  \hline
   $R_1$&$0.087$&$0.22$&$0.14$&$(0.03,0.5)$&$(0.04,0.4)$&$0.4\pm 0.2$\\
   $R_2$&$1.37$&$0.56$&$0.86$&$(0.4,2.2)$&$(0.3,3.8)$&$>21\pm 11$\\
   $R_3$&$10.24$&$4.76$&$6.88$&$(4.0,17.5)$&$(3.9,16.2)$&$13\pm 3$\\
   $R_4$&$1.76$&$0.59$&$1.13$&$(0.6,2.2)$&$(0.8,1.5)$&$1.1\pm 0.2$\\
   $R_5$&$0.85$&$0.38$&$0.57$&$(0.3,1.1)$&$(0.3,1.5)$&-\\
   $R_6$&$0.010$&$0.08$&$0.034$&$(0.01,0.1)$&$(0.001,0.1)$&-\\
   $R_7$&$0.84$&$1.81$&$1.24$&$(0.8,2.5)$&$(0.5,2.4)$&$1.3\pm 0.4$\\
   \hline\hline
  \end{tabular}
 \end{center}
 \label{tab:ra}
\end{table}
For comparison, we also show the values of the ratios for $\beta=0$ and $1$, in the second and third column of this table. In both methods, (I) and (II), the $95$\% CI obtained are comparable.  For the sample size used here, the 0.5 quantiles obtained are almost identical to the values of the ratios for $\beta=0.32$ in both methods, indicating that the sample is large enough in the present problem, and we omit those in the table. 

The experimental values fall inside the $95$\% CI for the ratios for $i=1,3,4,7$, and also these come out similar to those of Table 7 in Ref. \cite{alberdai}, showing a good agreement between theory and experiment, and also with the theory used in Ref. \cite{alberdai}. For the ratio $R_2$,  we obtain a value below the experimental lower  limit, for which we are inclined to blame the unprecise determination of the partial decay rate $\Gamma[J/\psi\to \omega f'_2(1525)]$.

The results obtained here for the ratios $i=1,3,4$ support the molecular VV nature of the tensor resonances, $f_2'(1270)$, $f_2'(1525)$ and $K^*_2(1430)$, investigated in Refs. \cite{raquel,gengvec}. The new ratio $R_7$ studied here is in perfect agreement with experiment, also  supporting the vector-vector nature of the scalar $f_0(1710)$. 

Although, in principle, different values of $\beta$ rather than that obtained in the fit, $0.32$, are possible, since the ratios obtained in the second and third columns are close or inside the $95$ \% CI given, $\beta=1$ seems to be an unlikely situation for $R_4$, according to the theory, because it is out of the intervals (I) and (II). On the contrary, the experimental value for $R_3$ seems to favor small values of $\beta$. However, these more extreme situations cannot be rejected with total confidence in the study done here, which favors an admixture of both terms, $\langle VVV\rangle$ and $\langle VV\rangle\langle V\rangle$. In any case, the results shown here for the four ratios $i=1,3,4,7$ provide further support to the molecular interpretation of the $f_2(1270),f_2'(1525),\bar{K}^*_2(1430)$ and $f_0(1710)$ of Refs. \cite{raquel,gengvec}. The predictions for the ratios $i=5,6$ involving the scalar resonances $f_0(1370)$ and $f_0(1710)$ are also given in Table \ref{tab:ra}, where $R_6$ turns out to be very small, indicating that the $J/\psi\to \phi f_0(1370)$ decay is suppressed. This can be easily understood in terms of the dominant term in Eq. (2), since the pair of vectors that create the $f_0(1370)$ are the $K^* \bar{K}^*$ and $\phi\phi$, but not $\rho\rho$ which is the main building block of the $f_0(1370)$ \cite{raquel,gengvec}.
\section{Summary and outlook}

    We have correlated different decays of the $J/\psi$ meson, like $J/\psi \to \phi (\omega) f_0(1370) [f_0(1710)]$,  $J/\psi \to \phi (\omega) f_2(1270) [f'_2(1525)]$,  and $J/\psi \to K^{*0} \bar{K}^{*0}_2(1430)$. These decay processes have in common the production of a vector meson together with some $f_0$, $f_2$, $K_2$ resonances, which in previous works have been shown to be dynamically generated from the vector-vector interaction.  The assumptions made are very basic, starting from the $J/\psi$, being a $c \bar c$ state, is a singlet of SU(3), in the same way as a $s \bar s$ state is a singlet of isospin SU(2). The nature of these resonances as dynamically generated implies that the production process begins by the  $J/\psi$ decaying into three vectors, followed by the interaction of a VV pair that leads to the formation of these resoanances.  Based on information gathered from related reactions, we find that the two basic SU(3) invariant structures that can be made from three vector SU(3) matrices, and are relevant to these reactions,  are $\langle VVV\rangle$ and $\langle V\rangle \langle VV\rangle$, with the first structure being dominant. We determine seven independent ratios that can be made with these decay rates and fit the strength of the $\langle V\rangle \langle VV\rangle$ structure versus the $\langle VVV\rangle$ one. We find indeed a relative strength of this term of 0.3. By means of this only parameter we can obtain good results compared to experiment for four ratios. Two extra ratios have no data to compare, and hence are genuine predictions of the theory, and one ratio disagrees with the data, for which tentatively we blame the lack of experimental information on the $J/\psi \to \omega f'_2(1525)$ decay, with only upper bound known. While certainly future measurements of the magnitudes involved in this ratio are most welcome, the overall agreement found with most data, with rates that are quite different to each other, speaks much in favor of the picture where these resonances are dynamically generated. The measurement of the rates needed to determine the two predictions for which there is no data at present would provide a further test for this idea and should be most welcome. 
\section*{Acknowledgments}
 L.R.D. and L.S.G acknowledge supports from the National Natural Science Foundation of China (Grant Nos. 11975009, 11575076, 11735003, 11975041). R.M. and E.O. acknowledge the hospitality of Beihang University where this work was initiated. This work is partly supported by the Talento Program of the Community of Madrid and the Complutense University of Madrid, under the project with Ref. 2018-T1/TIC-11167, and by the Spanish Ministerio de Economia y Competitividad and European FEDER funds under Contracts No. FIS2017-84038-C2-1-P B
and No. FIS2017-84038-C2-2-P B, and the Generalitat Valenciana in the program Prometeo II-2014/068, and
the project Severo Ochoa of IFIC, SEV-2014-0398 (EO).

\end{document}